\documentclass[prb,twocolumn]{revtex4-1}
\usepackage{amsfonts,amssymb,amsmath}
\usepackage{graphicx,psfrag}

\newcommand{\D}{\textrm{d}}

\begin{document}


\title{Friction force: from mechanics to thermodynamics}

\author{Christian Ferrari}
\address{Liceo di Locarno,\\ Via F. Chiesa 15, CH-6600 Locarno, Switzerland}
\author{Christian Gruber}
\address{Institut de Th\'eorie des Ph\'enom\`enes Physiques, Ecole Polytechnique F\'ed\'erale de Lausanne,\\ CH-1015 Lausanne, Switzerland}

\date{July 1$^{\text{st}}$, 2010. Accepted for European Journal of Physics}


\begin{abstract}
We study some mechanical problems in which a friction force is acting on the system. Using the fundamental concepts of state, time evolution and energy conservation we explain how to extend Newtonian mechanics to thermodynamics. We arrive at the two laws of thermodynamics and then apply them to investigate time evolution and heat transfer of some significant examples.
\end{abstract}


\maketitle


\section{Introduction}\label{sect1}

It is well known that thermodynamics was first developed by Carnot in 1824 to study the production and the transport of heat \cite{Carnot}. However since the axiomatic work of Carath\'eodory in 1909 on equilibrium thermodynamics \cite{carat}, also called thermostatics, one has very often considered thermodynamics as a theory restricted either to equilibrium states or transformations from one equilibrium state to another equilibrium state. Since 1960 St\"uckelberg has consistently presented thermodynamics in a way similar to mechanics, i.e. as a phenomenological theory with time evolution described by first order differential equations \cite{STS}. In his approach ``energy'' and ``entropy'' are extensive state functions introduced in the first and second law of thermodynamics. Furthermore the second law consists of two parts, one related to the time evolution of general systems and the other to the approach to equilibrium for isolated systems. By analogy with classical mechanics, where one starts with the study of a point particle to investigate then a finite number of interacting point particles and finally arrives at continuous systems, St\"uckelberg starts his study of thermodynamics by considering simple systems, which he calls ``element of system'', for which the state can be entirely described by only one non-mechanical, or thermal, variable -- the ``entropy'' whose existence is postulated in the second law -- together with the mechanical variables. He then considers general systems made of a finite or infinite number of interacting simple systems. In such an approach thermostatics appears as a special chapter of thermodynamics just as statics appears as a special chapter in textbooks on mechanics. Moreover it should be stressed that mechanics appears as a special case of thermodynamics where the mechanical observables are decoupled from the thermal ones.

In this paper we illustrate St\"uckelberg's approach with some simple mechanical systems submitted to friction force. Using the structural properties discussed below, and assuming the conservation of energy, we show how one is lead to extend mechanics to arrive at thermodynamics. In particular, following St\"uckelberg, we formulate the first and second law as time evolution described by first order differential equations. These examples will be used to illustrate the evolution of mechanical systems in the presence of friction taking into account the thermodynamical aspects, since it is well known that the temperature increases as soon as friction is introduced. Previous analysis in similar directions have been done \cite{A,BG,UB, CEM,S}. They differ from our approach in which we use observables defined at each instant of time and we use time dependent formalism to describe the time evolution. Moreover we explicitly introduce the entropy concept and arrive at the second law of thermodynamics.

It is well known that a classical physical system is characterized by its \emph{observables}, its \emph{state} and the \emph{time evolution}. These three concepts are structural, in the sense that they can be defined in many branches of physics. The observables are those physical quantities chosen by one observer to study the system. The state at time $t$ represents the information hold by the observer on the possible values of those observables at time $t$. The time evolution yields the state at time $t$ knowing the state at some initial time $t_0$ together with the action of the outside on the system at all times. In the following we consider only pure states, i.e. the state at time $t$ gives the precise value of each observable at this time.

Let us present the main ideas of our work. We assume that for every physical system there is an observable, called the energy, which is conserved. This means that the numerical value of this observable remains constant whenever the system is either isolated from the outside or submitted to passive forces only, i.e. forces which do not produce work, sometimes called zero-work force \cite{A}. Example of passive forces are those forces whose application point does not move, such as the force exerted on particles by an ideal fixed wall, or those forces which are always perpendicular to the velocity such as the force exerted by a magnetic field on an electric charge, or a perfect constraint on the motion. In other words this assumption implies that the only way to change the energy of the system is by an action from the outside.  In mechanics it follows from Newton's definition in the Principia that the only way the outside can act on the system and change its state is by means of ``force''. Looking at the system from a mechanical point of view, in particular assuming that the only action of the outside on the system is by means of force, it is immediately clear that mechanical energy is not a conserved observable when friction forces are present. Therefore to have an observable energy which is conserved we are forced to introduce a new form of energy, namely the internal energy, and the energy appears as the sum of the mechanical and the internal energy (recall that the mechanical energy is the sum of several contributions which appear in the form of kinetic and potential energies). Then one finds that to define the state of the system we need new non-mechanical, or thermodynamical, variables. Assuming the simplest case where it is sufficient to introduce only one thermodynamical variable, we take the entropy and arrive at the second law of thermodynamics: for an adiabatically closed system the entropy can never decrease. Now that we have reached the conclusion that thermodynamical variables are needed to define the state we are then led to conclude that it is possible for the outside to act on those variables to change the state of the system, in particular its energy, without any force; this leads to the concept of heat power, or quantity of heat delivered to the system from the outside per unit of time: it is positive if the system receives heat and negative if the system gives heat to the outside. Of course in more general cases it will not be the only new contribution to the energy; for example if we consider a car accelerating on an horizontal road it will be necessary to introduce a new form of energy, the chemical energy, new state variables, the number of moles of each elements, and new actions of the outside on the system, i.e. the transfer of matter from the outside; in other examples we will be led to introduce other forms of energy such as nuclear energy and relativistic rest energy.

In section~\ref{sect2} we present a simple frictionless mechanical problem. In section~\ref{sect3} we consider the more general case in which friction is present and we extend the mechanical scenario to a more general one leading to thermodynamics. Section~\ref{sect4} is devoted to significant example that can be easily calculated and in which we can get an insight about energy transfer between two bodies in the presence of friction forces. Moreover we calculate the time evolution of the temperature, of the entropy, and the heat transfer per unit of time between two bodies.

This work is inspired from \cite{CG1} and a complete description of this formalism can be found in the original work of St\"uckelberg and Scheurer \cite{STS}. It is intended for undergraduates with some knowledge of equilibrium thermodynamics as well as for all those believing that thermodynamics is restricted to equilibrium states.
As we have personally experienced, the explicit connection between mechanics and thermodynamics used in the paper can provide a significant improvement in the teaching of thermodynamics in the time dependent approach with frictional forces.\\ Moreover, this approach clearly shows that the first law of thermodynamics is valid even when work is done against friction forces.


\section{The frictionless problem}\label{sect2}

The system $\Sigma$ is the union of a solid $\Sigma_1$ moving on a solid $\Sigma_2$, which is maintained fixed by means of some applied force, as represented in figure~\ref{fig2td}. This force is thus passive and has no effect on the energy of the system (although it has of course the effect that momentum is not constant).
The two subsystems interact by means of some internal conservative force $\vec F^{1\to 2}=-\vec F^{2\to 1}$ (which is represented by a massless spring in figure~\ref{fig2td}) and we assume that there is no friction force between the two solids, or with air. The mechanical state of $\Sigma_1$ and $\Sigma_2$ is given by their position $\vec x_i$ and their momentum $\vec p_i$ ($i=1,2$). Since $\Sigma_2$ is at rest its mechanical state $(\vec x_2, \vec p_2=\vec 0)$ does not change in time. Thus the state of the total system $\Sigma$ can be described by $(\vec x_1,\vec p_1)\equiv (\vec x,\vec p)$ of the moving solid.\\ To be explicit let us consider the special, but not restrictive, case where the force $\vec F^{2\to 1}$ is given by an elastic force of the form $\vec F_{\rm el}=-k\vec x$ where $k$ is a positive constant.

Next we describe the time evolution, which follows from Newton's second law and his definition of momentum, $\vec p=m\vec v$, where $m$ is the mass of $\Sigma_1$. Newton's equations are
\begin{subequations}
\begin{align}
\frac{\D \vec x}{\D t}&=\frac{\vec p}{m} \label{q1}\\
\frac{\D \vec p}{\D t}&=\vec F_{\rm el} \label{q2}\ .
\end{align}
\end{subequations}
This is a system of first order differential equations that allows to calculate the state at every instant $t$ knowing the state $(\vec x_0,\vec p_0)$ at some instant $t_0$ (the initial data) and the force $\vec F_{\rm el}$: it is the determinism principle. Note that $\vec x$ and $\vec p$ are two independent variables which vary periodically.

We finally switch to some energy consideration. From \eqref{q1} and \eqref{q2} we get
\begin{align}
\frac{\D E^{\rm kin}_1}{\D t}&=\vec F_{\rm el}\cdot \vec v
\end{align}
where  $E^{\rm kin}_1=\frac{1}{2}m v^2$ with $v=|\vec v|$ and since $\Sigma_2$ is fixed $\frac{\D E^{\rm kin}_2}{\D t}=0$. From $\vec F_{\rm el}=-k\vec x=-\nabla E^{\rm pot}$, it follows that $\vec F_{\rm el}$ is conservative, and we have
\begin{equation}
 \frac{\D E^{\rm kin}_1}{\D t}=-\nabla E^{\rm pot}\cdot \vec v=-\frac{\D E^{\rm pot}}{\D t}\ .
\end{equation}
We thus have conservation of the mechanical energy $E^{\rm mec}=E^{\rm kin}_1+E^{\rm kin}_2+E^{\rm pot}$, i.e.
\begin{equation}
\frac{\D E^{\rm mec}}{\D t}=0\ .
\end{equation}


\section{Friction, internal energy and entropy}\label{sect3}

If all the active forces (i.e. those which produce work) are conservative the mechanical energy, i.e. kinetic plus potential energy, is conserved.
This means that the mechanical energy can be modified only by some external action on the system. This \emph{conservation of mechanical energy} is a fundamental property since, using Noether's theorem \cite{Gold}, one can show that it is associated with the temporal invariance of physical laws.
Moreover using the fact that at the microscopic level all forces are conservative, we want to assume that at the macroscopic level there always exists an extensive state function, the energy $E$, which is a conserved quantity. Therefore, assuming that the only way the outside can act on the system is by means of forces, we should have
\begin{equation}\label{g1}
  \frac{\D E}{\D t}=P_W^{\rm ext}
\end{equation}
where $P_W^{\rm ext}$ is the power developed by the external forces acting on the system.

The assumption \eqref{g1} is the first law of thermodynamics for adiabatically closed systems, i.e. those systems where the only external actions on the system are by means of forces.

\begin{figure}[h]
 \begin{center}
\psfrag{S}[c][][1.2]{$\Sigma$}
\psfrag{1}[c][][1]{$\Sigma_2$}
\psfrag{2}[c][][1]{$\Sigma_1$}
\psfrag{F}[c][][1]{$\vec F_{\rm ext}$}
\psfrag{x}[c][][1]{$\vec x_0$}
\psfrag{x1}[c][][1]{$\vec x_1$}
\psfrag{p}[c][][1]{$\vec p_0=\vec 0$}
\psfrag{p1}[c][][1]{$\vec p_1=\vec 0$}
\psfrag{O}[c][][1]{$O$}
\includegraphics[height=3.3cm]{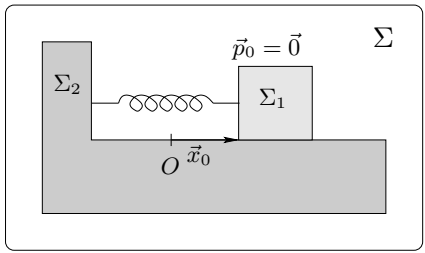}
\caption{The solid $\Sigma_1$ moves on a fixed solid $\Sigma_2$ subjected to a friction force and to an elastic force. The initial mechanical state of $\Sigma$ is $(\vec x_0\not =\vec 0,\vec p_0=\vec 0)$.}
\label{fig2td}
 \end{center}
\end{figure}

Let us consider the situation represented in figure~\ref{fig2td}.

\noindent The total system $\Sigma$ is identical to that studied in the previous section, but now the solid $\Sigma_1$ is also subjected to the dissipative force $\vec F_{\rm fr}$ arising from the friction on the surface of $\Sigma_2$. After $\Sigma_1$ is released from a non-equilibrium state $(\vec x_0\not =\vec 0,\vec p_0=\vec 0)$, it will have damped oscillations until it reaches an equilibrium state $(\vec x_1,\vec p_1=\vec 0)$.

We can assume, as usual for solid friction, that this friction force is given by
\begin{equation}\label{solidfr}
\begin{cases}
  \vec F_{\rm fr}=-\lambda \hat v & \text{if $v\not =0$} \qquad \left(\hat v=\frac{\vec v}{|\vec v|}\right) \\
  |\vec F_{\rm fr}|\leq F_{\rm fr,max} & \text{if $v =0$}
\end{cases}
\end{equation}
where $\lambda=\lambda(\vec x,\vec p)$ is strictly positive.

The evolution is described by Newton's equations
\begin{subequations}
\begin{align}
\frac{\D \vec x}{\D t}&=\frac{\vec p}{m}\label{q1b}\\
\frac{\D \vec p}{\D t}&=\vec F_{\rm el}+\vec F_{\rm fr}\label{q2b}
\end{align}
\end{subequations}
if $\vec p\not =\vec 0$ or $\vec p=\vec 0$ and $|\vec F_{\rm el}|>F_{\rm fr,max}$, otherwise $\frac{\D \vec p}{\D t}=\vec 0$.
Therefore
\begin{equation}\label{1tdm}
\frac{\D E^{\rm mec}}{\D t}=P^{({\rm nc})}
\end{equation}
where $P^{({\rm nc})}=\vec F_{\rm fr}\cdot \vec v$, the power associated with the non conservative friction force, is strictly negative if $v$ is non zero. Therefore
\begin{equation}\label{pnc}
  P^{({\rm nc})}=-\lambda v\leq 0
\end{equation}
and the mechanical energy is not conserved. In fact using \eqref{1tdm} and \eqref{pnc}, the mechanical energy in the equilibrium state $(\vec x_1,\vec p_1=\vec 0)$ will be strictly smaller than in the initial state $(\vec x_0\not =\vec 0,\vec p_0=\vec 0)$.

For our assumption \eqref{g1} to be valid (with $P_W^{\rm ext}=0$), we have to assume that there exists some state function $U$, called \emph{internal energy} of the whole system $\Sigma$, such that if $v\not =0$
\begin{equation}\label{introU}
\frac{\D U}{\D t} = -P^{({\rm nc})} \ .
\end{equation}
Only in this case will we be able to conclude that there is conservation of energy $E$ defined by
 \begin{equation}
 E=E^{\rm mec} + U\ .
\end{equation}

\noindent If such a function exists we shall be able to conclude that in the equilibrium state ($\vec x_1, \vec p_1=0$), part of the initial potential energy has been converted into internal energy, $U_1-U_0=E_0^{\rm pot}-E_1^{\rm pot}$, but the total energy is constant $E_1=E_0$.

\begin{figure}[h]
 \begin{center}
\psfrag{S}[c][][1.2]{$\Sigma$}
\psfrag{S1}[c][][1]{$\Sigma_1$}
\psfrag{S2}[c][][1]{$\Sigma_2$}
\psfrag{F}[c][][1]{$\vec F_{\rm ext}$}
\psfrag{x0}[c][][1]{$\vec x_0$}
\psfrag{x1}[c][][1]{$\vec x_1$}
\psfrag{p0}[c][][1]{$\vec p_0=\vec 0$}
\psfrag{p1}[c][][1]{$\vec p_1=\vec 0$}
\psfrag{O}[c][][1]{$O$}
\includegraphics[height=3.2cm]{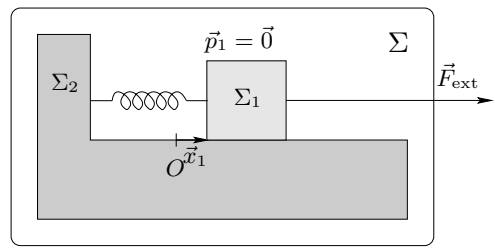}
\caption{In a second step, after the system has reached the mechanical equilibrium $(\vec x_1,\vec p_1=\vec 0)$, an external force $\vec F_{\rm ext}$ is applied to the system $\Sigma_1$ to recover the initial mechanical configuration $(\vec x_0\not =\vec 0,\vec p_0=\vec 0)$.}
\label{fig3td}
 \end{center}
\end{figure}

In a second step, see figure~\ref{fig3td}, we want to restore the system $\Sigma$ in its initial mechanical state $(\vec x_0\not =\vec 0,\vec p_0=\vec 0)$. To do so we have to apply an external force $\vec F_{\rm ext}$.

The evolution of the mechanical state is given by Newton's equations
\begin{subequations}
\begin{align}
\frac{\D \vec x}{\D t}&=\frac{\vec p}{m}\label{q1c}\\
\frac{\D \vec p}{\D t}&=\vec F_{\rm el}+\vec F_{\rm fr} + \vec F_{\rm ext}\label{q2c}
\end{align}
\end{subequations}
which implies
\begin{equation}
  \frac{\D E^{\rm mec}}{\D t}=P^{({\rm nc})}+\vec F_{\rm ext}\cdot \vec v\ .
\end{equation}
Therefore with our assumption \eqref{introU} we obtain the desired result
\begin{equation}\label{1aa}
  \frac{\D E}{\D t}=P_W^{\rm ext}\ ,
\end{equation}
with
\begin{equation}\label{defPW}
P_W^{\rm ext}=\vec F_{\rm ext}\cdot \vec v
\end{equation}
i.e. the basic assumption \eqref{g1} is satisfied.

However, since $E_1 = E_0$, we note that the work done by the external force,
\begin{equation}
 W^{\rm ext}=\int_{t_1}^{t_2} P_W^{\rm ext}(t) \D t=E_2-E_1=E_2-E_0=U_2-U_0
\end{equation}
is always strictly positive and thus $U_2>U_0$.\\
In conclusion, if we want to assume that $U$, and thus $E$, are state functions, the fact that the original and final mechanical states are identical
 but $U_2\not =U_0$ forces us to introduce new non-mechanical state variables to distinguish the final state from the original one. Therefore the state of the system $\Sigma$ must be described by $(\vec x,\vec p)$ together with some non-mechanical state variables.

The simplest case is the one where it is sufficient to introduce just one non-mechanical variable. We could chose the temperature since we observe that the system becomes warmer, or the internal energy which we were led to introduce. We prefer to introduce a new observable $S$, called the entropy, which will be specified below (see \eqref{tempass}). In this simplest case the state of the system $\Sigma$ is described by $(\vec x,\vec p,S)$. Explicitly the dependence of the energy on the state variables is
\begin{equation}\label{enf}
E(\vec x,\vec p, S)=E^{\rm kin}(\vec p,S)+E^{\rm pot}(\vec x,S)+U(S) \ .
\end{equation}
We shall choose the entropy $S$ in such a way that the state function
\begin{equation}\label{tempass}
  T=\frac{\partial E}{\partial S}
\end{equation}
can be identified with the \emph{absolute temperature} of $\Sigma$, since in an equilibrium state the temperature is defined by \eqref{tempass}. The internal energy $U$ must be a function of $S$, while the mass, the potential energy, and the friction coefficient could be functions of $S$, $m=m(S)$, $E^{\rm pot}=\frac{1}{2}k(S)x^2$, and $\lambda=\lambda(\vec x,\vec p,S)$.

We now want to investigate the time evolution of the state $(\vec x,\vec p,S)$. From \eqref{enf} we have
\begin{equation}
\frac{\D E}{\D t}=\nabla_{\vec p} E\frac{\D \vec p}{\D t}+ \nabla E\frac{\D \vec x}{\D t}+ \frac{\partial E}{\partial S}\frac{\D S}{\D t}\ .
\end{equation}
Using Newton's equations \eqref{q1c} and \eqref{q2c} for $\frac{\D \vec x}{\D t}$ and $\frac{\D \vec p}{\D t}$, together with 
\begin{equation}
\nabla_{\vec p} E=\frac{\vec p}{m},\quad \nabla E=-\vec F_{\rm el}, \quad \frac{\partial E}{\partial S}=T
\end{equation}
we obtain
\begin{equation}\label{www}
\frac{\D E}{\D t}=\frac{\vec p}{m}\cdot (\vec F_{\rm el}+\vec F_{\rm fr}+\vec F_{\rm ext}) - \vec F_{\rm el}\cdot \frac{\vec p}{m} +T \frac{\D S}{\D t} \ .
\end{equation}

\noindent Therefore, from  \eqref{1aa} and \eqref{www}, we have
\begin{equation}\label{qwe}
  \frac{\D E}{\D t}=P_W^{\rm ext}=\frac{\vec p}{m}\cdot (\vec F_{\rm fr}+\vec F_{\rm ext})+ T \frac{\D S}{\D t}\, ,
\end{equation}
and with the definition $P_W^{\rm ext}=\vec F_{\rm ext}\cdot \vec v$ we obtain
\begin{equation}\label{2td1}
\frac{\D S}{\D t}=-\frac{1}{T}\vec F_{\rm fr}\cdot \frac{\vec p}{m}=\frac{1}{T}\lambda \frac{p}{m}\geq 0 \ .
\end{equation}
where $p=|\vec p|$.
This last equation is the second law of thermodynamics for adiabatically closed systems. The non-negative state function
\begin{equation}\label{prodSI}
  I(\vec x,\vec p,S)=\frac{1}{T}\lambda \frac{p}{m}
\end{equation}
is called \emph{entropy production} and characterizes irreversible processes.

The time evolution of $\Sigma$ is then given by the ODE
\begin{subequations}
\begin{align}
\frac{\D \vec x}{\D t}&=\frac{\vec p}{m}\label{mec1} \\
\frac{\D \vec p}{\D t}&=-k\vec x -\frac{\lambda}{m} \hat p+\vec F_{\rm ext} \label{mec2}\\
\frac{\D S}{\D t}&=\frac{1}{T} \frac{\lambda}{m}p \label{td}
\end{align}
\end{subequations}
if $\vec p\not =\vec 0$ or $\vec p=\vec 0$ and $|-k\vec x(t)+\vec F_{\rm ext}|>F_{\rm fr,max}$ (in the latter case $-\frac{\lambda}{m} \hat p$ has to be replaced by the static solid friction force), otherwise $\frac{\D \vec p}{\D t}=\vec 0$ and thus $\vec p=\vec 0$.

The friction coefficient $\lambda=\lambda(\vec x,\vec p, S)$, the spring constant $k=k(S)$ and the inertial mass \mbox{$m=m(S)$} should be defined from experiments.

\noindent We observe that the two equations \eqref{mec1} and \eqref{mec2}, that characterize the mechanical problem, are now coupled with the new one \eqref{td}, which is related to thermodynamics. In general $k$, $\lambda$, $m$ do not depend very strongly on $S$ (or equivalently on temperature). If this dependence can be neglected we can solve \eqref{mec1} and \eqref{mec2} independently of the thermodynamic equation \eqref{td}. This justifies the study of friction in mechanics without taking into account the thermodynamical aspect.

Up to now, the only external action on the system was by means of external forces acting on the state variable $\vec x$ and expressed by $P_{W}^{\rm ext}$ (see \eqref{defPW}).
Having introduced a new variable $S$ we can now introduce a new type of external action which can modify $S$ and thus the energy without changing $\vec x$ or $\vec p$. This action can be considered as acting on the microscopic variables which have disappeared at the macroscopical level to be replaced by the single variable $S$.
We express this external action by $P_Q^{\rm ext}$, the \emph{heat power}, which is the heat transferred from the outside per unit time. We now have
\begin{equation}\label{q}
\frac{\D E}{\D t}=P_W^{\rm ext}+P_Q^{\rm ext}
\end{equation}
which is the \emph{first law of thermodynamics} for closed systems (i.e. systems that do not exchange matter).

\noindent To discover the effect of $P_Q^{\rm ext}$ on the entropy we observe that from \eqref{www}, \eqref{prodSI} and \eqref{defPW} we obtain
\begin{equation}
P_Q^{\rm ext}=-\lambda\frac{p}{m}+T\frac{\D S}{\D t}
\end{equation}
that is from \eqref{prodSI}
\begin{equation}\label{secondalegge}
  \frac{\D S}{\D t}=I+\frac{1}{T}P_Q^{\rm ext} \qquad I\geq 0\ .
\end{equation}
This last equation is the \emph{second law of thermodynamics} for closed system,  with $I$, the entropy production, which is a non negative state function \cite{STS}.

To conclude this section we remark that the non-negative term $-\vec F_{\rm fr}\cdot \vec v=\frac{\lambda}{m}p$ is usually called dissipative power and in this example the entropy production is
 $ I=\frac{1}{T}P^{\rm dis}\ .$


\section{Example}\label{sect4}

\begin{figure}[h]
 \begin{center}
\psfrag{S2}[c][][1]{$\Sigma_2$}
\psfrag{S1}[c][][1]{$\Sigma_1$}
\psfrag{Fa}[c][][1]{$\vec F_{\rm ext}$}
\includegraphics[height=3.5cm]{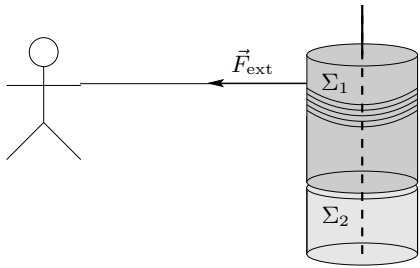}
\caption{The solid $\Sigma_1$ turns on a solid $\Sigma_2$ and is subjected to a friction force, an external force is applied to $\Sigma_1$ such that it turns with constant angular velocity.}\label{ex3a}
 \end{center}
\end{figure}

\noindent  We consider the system described by figure~\ref{ex3a}, where the cylinder $\Sigma_2$ is fixed and the cylinder $\Sigma_1$ can rotate around its axes which is vertical. Passive external forces are applied to maintain fixed $\Sigma_2$ and the axis of $\Sigma_1$. A constant horizontal external force $\vec F_{\rm ext}$ acts on the cylinder $\Sigma_1$ and we shall discuss the stationary mechanical state when the angular velocity $\vec \omega$ is constant. The radius of the cylinders is $R$.

\noindent We assume that experiments have shown that the masses and the friction torque $\lambda$ are positive constants and the internal energy of the solid $\Sigma_i$ is given by $U_i=3n_iRT_i$, where $n_i$ is the number of moles of $\Sigma_i$  ($i=1,2$).
Following the discussion of section~\ref{sect3} we assume that the state of the system is entirely described by only one thermodynamical variable, which implies that the state is described by only one temperature $T$, i.e. we are thus assuming that both cylinders are at all times at the same temperature $T = T_1 = T_2$. (The more general case where the two cylinders have different temperatures is discussed in \ref{appD}).  Since the energy is an extensive variable, the internal energy of the whole system $\Sigma$ is thus $U=3(n_1+n_2)RT$.

Under the above conditions we have for the whole system $\Sigma$ conservation of angular momentum
\begin{equation}
0=F_{\rm ext} R - \lambda
\end{equation}
and the first law of thermodynamics \eqref{q} writes as
\begin{equation}\label{Exo1}
\frac{\D E}{\D t}=P_W^{\rm ext}\ ,
\end{equation}
where $P_W^{\rm ext}=\vec F_{\rm ext}\cdot \vec v=\lambda \omega$ where $|\vec v|=\omega R$. Since the angular velocity of $\Sigma_1$ is constant and $\Sigma_2$ is fixed, $\frac{\D E^{\rm mec}}{\D t}=0$ and \eqref{Exo1} imply
\begin{equation}
\frac{\D U}{\D t}=\lambda \omega\ ,
\end{equation}
therefore
\begin{equation}\label{tempe}
3(n_1+n_2)R\frac{\D T}{\D t}=\lambda \omega
\end{equation}
Note that we can apply the first law of thermodynamics \eqref{q} to $\Sigma_1$ and $\Sigma_2$ separately. For $\Sigma_1$ we have
\begin{equation}
  \frac{\D E_1}{\D t}=P_Q^{2\to 1}+ \underbrace{P_W^{\rm ext}+P_W^{2\to 1}}_{=0}
\end{equation}
which gives
\begin{equation}
\frac{\D U_1}{\D t}=P_Q^{2\to 1}
\end{equation}
 and thus with \eqref{tempe}
\begin{equation}
  P_Q^{2\to 1}=3n_1R\frac{\D T}{\D t}=\frac{n_1}{n_1+n_2}\lambda \omega\ .
\end{equation}
For $\Sigma_2$ we have
\begin{equation}
  \frac{\D E_2}{\D t}=P_Q^{1\to 2}+ P_W^{1\to 2}
\end{equation}
which gives
\begin{equation}
\frac{\D U_2}{\D t}=P_Q^{1\to 2}+P_W^{1\to 2}=P_Q^{1\to 2}+\lambda \omega \ .
\end{equation}
Therefore, from $\frac{\D E}{\D t}=\frac{\D U_1}{\D t}+\frac{\D U_2}{\D t}=P_W^{\rm ext}$ and $P_W^{\rm ext}=\lambda \omega$ we conclude that
\begin{equation}\label{55}
  P_Q^{1\to 2}=-P_Q^{2\to 1}=-\frac{n_1}{n_1+n_2}\lambda \omega\ .
\end{equation}
Finally,
\begin{equation}\label{56}
\frac{\D E_1}{\D t}=\frac{n_1}{n_1+n_2}\lambda \omega \qquad \text{and} \qquad  \frac{\D E_2}{\D t}=\frac{n_2}{n_1+n_2}\lambda \omega\ .
\end{equation}

Let us note that \eqref{55} and \eqref{56}, together with \eqref{tempass} imply
\begin{eqnarray}
 \frac{\D S_1}{\D t} &=& \frac{1}{T}  P_Q^{2\to 1} \\
 \frac{\D S_2}{\D t}  &=&  I + \frac{1}{T}  P^{1\to 2}_Q \qquad \text{with} \qquad I = \frac{\lambda \omega }{T}> 0\\
 \frac{\D S}{\D t} &=& I
\end{eqnarray}
which are the expression of the second law \eqref{secondalegge} applied to the subsystems $\Sigma_1$, $\Sigma_2$  and the whole system $\Sigma$ which is adiabatically closed.

We now address the question of time evolution. From \eqref{tempe} the temperature time evolution is given by
\begin{equation}
T(t)=T_0+\frac{\lambda \omega }{3(n_1+n_2)R}t\ .
\end{equation}
Since by definition $\frac{\partial E}{\partial S} =T$ we have, in this example,
\begin{equation}
 T =\frac{\partial E}{\partial S}= \frac{\D U}{\D S} = 3(n_1+n_2) R \frac{\D T}{\D S}
\end{equation}
and thus the entropy is related to the temperature by the function
 \begin{equation}\label{entropo}
 \begin{aligned}
 & S=S_0+3\left(n_1+n_2\right)R\ln \frac{T}{T_0}\\
 &=S_0+3\left(n_1+n_2\right)R\ln \frac{U}{U_0}\ .
 \end{aligned}
\end{equation}
We finally obtain the time evolution for the entropy of the system
\begin{equation}
\begin{aligned}
S(t)&=S_0+3(n_1+n_2)R\ln \frac{T(t)}{T_0}\\
&=S_0+3(n_1+n_2)R\ln \Big(1+\frac{\lambda \omega}{3(n_1+n_2)RT_0}t\Big)\ .
\end{aligned}
\end{equation}
We observe that the total entropy $S(t)$ is an increasing function of time. Since the system $\Sigma$ is adiabatically closed \mbox{$(P_Q^{\rm ext}=0)$} this is once more the second law of thermodynamics.

To gain some more insight into this example we consider $\Sigma_2$ as $\tilde\Sigma_2\cup \Sigma_3$ where the subsystem $\Sigma_3$ is defined by the surface of $\Sigma_2$ whose height is $\varepsilon \ll 1$.

\noindent We now apply the first law of thermodynamics to these three subsystems
\begin{eqnarray}
\frac{\D U_1}{\D t}&=&\underbrace{P_W^{\rm ext}+P_W^{3\to 1}}_{=0}+P_Q^{3\to 1},\\
\frac{\D \tilde U_2}{\D t}&=&P_Q^{3\to 2},\\
\frac{\D U_3}{\D t}&=&P_W^{1\to 3}+P_Q^{1\to 3}+P_Q^{2\to 3},
\end{eqnarray}
where we used the fact no work is done by $\Sigma_3$ on $\tilde \Sigma_2$ and thus $P_W^{3\to 2}=P_W^{2\to 3}=0$.
But, using Newton's third law, $P_W^{1\to 3}=\lambda \omega=P_W^{\rm ext}$. Since  $U_3$ is proportional to $\varepsilon$, in the limit $\varepsilon\to 0$ we have $U_3\to 0$ and $\tilde U_2\to U_2$, and thus in this limit
\begin{equation}\label{cvb}
0=P_W^{1\to 3}+P_Q^{1\to 3}+P_Q^{2\to 3}\ .
\end{equation}
Moreover the equation
\begin{equation}
\begin{aligned}
   P_W^{\rm ext}&= \frac{\D E}{\D t}=\frac{\D U_1}{\D t}+\frac{\D \tilde U_2}{\D t}+\frac{\D U_3}{\D t}\\
  &=P_Q^{3\to 1}+P_Q^{3\to 2}+P_W^{1\to 3}+P_Q^{1\to 3}+P_Q^{2\to 3}
\end{aligned}
\end{equation}
together with \eqref{cvb} gives finally
\begin{equation}\label{wer}
  P_W^{\rm ext}=P_Q^{3\to 1}+P_Q^{3\to 2}\ .
\end{equation}

\noindent We see that the external work power given to the system appears as heat power at the interface, which is then distributed to the two subsystems $\Sigma_1$ and $\Sigma_2$ in proportion to their number of moles
\begin{equation}
P_Q^{3\to i}=3n_iR\frac{\D T}{\D t}=\frac{n_i}{n_1+n_2}P_W^{\rm ext} \qquad (i=1,2)\ .
\end{equation}

Others examples are given in the appendices.


\section{Conclusions}\label{sect5}

Applying the idea of state and time evolution, together with the basic  assumption of energy conservation, to simple examples we have shown that one is naturally led from mechanics to thermodynamics. This can be used to  study mechanical problems with friction from a more general point of  view. In particular the second law, the production of entropy, and the  question of heat transfer can be discussed in the time dependent thermodynamic approach such as developed by St\"uckelberg \cite{STS}.

Moreover, this formalism can be applied to many topics in thermodynamics, in particular to those that cannot be analyzed with the equilibrium thermodynamics \cite{CG3}; as we show in the appendices with the solid friction example the second law of thermostatics is not always satisfied.


\begin{acknowledgments}
The authors would like to thanks Professor Patrik Ferrari and Professor Philippe Martin for stimulating discussions and valuable comments, and the anonymous referee of Eur. J. Phys. for a careful review and constructive comments.
\end{acknowledgments}


\appendix

\section{Viscous friction}\label{secA}

\begin{figure}[h]
 \begin{center}
\psfrag{S2}[c][][1]{$\Sigma_2$}
\psfrag{S1}[c][][1]{$\Sigma_1$}
\psfrag{Fa}[c][][1]{$\vec F_{\rm fr}$}
\psfrag{Fel}[c][][1]{$\vec F_{\rm el}$}
\includegraphics[height=1.8cm]{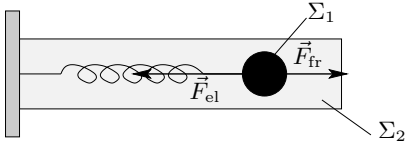}
\caption{The solid $\Sigma_1$ moves in an ideal gas $\Sigma_2$ and is subjected to a viscous friction force and to an elastic force.}\label{ex2}
 \end{center}
\end{figure}

We consider the system illustrated in figure~\ref{ex2}. The solid $\Sigma_1$, whose mass is $m$, is subjected to a viscous friction force $\vec F_{\rm fr}=-\lambda \vec v$ and an elastic force $\vec F_{\rm el}=-k\vec x$; where we assume that $m$, $\lambda$ and $k$ are positive constants.
Moreover the internal energy of $\Sigma_1$ is supposed to be given by $U_1=3n_1RT_1$ with $n_1$ its number of moles. To simplify the analysis $\Sigma_2$ is modeled as a monoatomic ideal gas with internal energy $U_2=\frac{3}{2}n_2RT_2$ and fixed volume and mole number $n_2$.
The only action of the outside on the system is the passive force maintaining fixed the cylinder containing the ideal gas.
Since we assume that the state is entirely described by only one thermodynamical variable $S$, together with $\vec x$ and $\vec p$ of $\Sigma_1$,
both the solid and the fluid are at all time at the same temperature $T=T_1=T_2$.
Moreover since energy is an extensive observable the system's energy is thus $E=\frac{1}{2}mv^2+\frac{1}{2}kx^2+(3n_1+\frac{3}{2}n_2)RT$.

The initial condition for the mechanical state is $\vec x_0\not =\vec 0$ and $\vec p_0=m\vec v_0\not =\vec 0$, we also assume that $\kappa=\frac{\lambda}{2m}$ and $\omega_0=\sqrt{\frac{k}{m}}$ are such that $\kappa<\omega_0$.

The ODE giving the time evolution of the state $(\vec x,\vec p,S)$ is
\begin{subequations}
\begin{align}
\frac{\D \vec x}{\D t}&=\frac{\vec p}{m}\label{mec1v} \\
\frac{\D \vec p}{\D t}&=-k\vec x -\frac{\lambda}{m} \vec p \label{mec2v}\\
\frac{\D S}{\D t}&=\frac{1}{T}\lambda \left(\frac{p}{m}\right)^2 \label{tdv}
\end{align}
\end{subequations}
where $p=|\vec p|$.

Since $m,\lambda$ and $k$ are constant we can solve Newton's equations \eqref{mec1v} and \eqref{mec2v}. The time evolution of the position is thus given by the damped oscillations (\mbox{$\omega=\sqrt{\omega_0^2-\kappa^2}$})
\begin{equation}
x(t)=\text{e}^{-\kappa t} \left[ x_0\cos (\omega t)+\frac{v_0+\kappa x_0}{\omega}\sin(\omega t)\right]
\end{equation}
while the time evolution of the momentum is given by
\begin{equation}
p(t)=m\text{e}^{-\kappa t}\left[v_0 \cos(\omega t)
- \left(\kappa{\frac { v_0+\kappa x_0  }{\omega}} +x_0\omega \right)\sin ( \omega t )  \right] \ .
\end{equation}

As in section~\ref{sect4}, using the definition $\frac{\partial E}{\partial S} =T$ we have, in this example,
\begin{equation}
 T =\frac{\partial E}{\partial S}= \frac{\D U}{\D S} = 3(n_1+\tfrac{1}{2}n_2) R \frac{\D T}{\D S}
\end{equation}
and thus the entropy is related to the temperature by the function
 \begin{equation}
\begin{aligned}
&  S=S_0+3\left(n_1+\tfrac{1}{2}n_2\right)R\ln \frac{T}{T_0}\\
&=S_0+3\left(n_1+\tfrac{1}{2}n_2\right)R\ln \frac{U}{U_0}
  \end{aligned}
\end{equation}

\noindent Therefore to obtain the evolution of the entropy $S(t)$, we shall first find the evolution of the temperature $T(t)$. Under the above conditions and \eqref{tdv} we have
\begin{equation}
\frac{\D U}{\D t}=3\left(n_1+\tfrac{1}{2}n_2\right)R\frac{\D T}{\D t}=\frac{\D U}{\D S}\frac{\D S}{\D t} = T \frac{\D S}{\D t} =\lambda \left(\frac{p}{m}\right)^2
\end{equation}
from which we obtain the time evolution for the temperature
\begin{equation}
T(t)=T_0 +\frac{1}{3\left(n_1+\tfrac{1}{2}n_2\right)R} f(t)
\end{equation}
with
\begin{equation}
\begin{aligned}
&f(t)=\left(\tfrac{1}{2}mv_0^2+\tfrac{1}{2}kx_0^2\right)-\frac{1}{2(4km-\lambda^2)}\text{e}^{-\frac{\lambda}{m} t}\\
&\times\Big[4km(mv_0^2+\lambda v_0x_0+kx_0^2) \\
&- \lambda(v_0^2\lambda m+4 v_0mkx_0+\lambda kx_0^2 )\cos(2 \omega t)\\
&-\lambda (mv_0^2-kx_0^2)(4km-\lambda^2)^{1/2}\sin(2 \omega t)\Big]\ .
\end{aligned}
\end{equation}
which gives the entropy time evolution
\begin{equation}
    S(t)=S_0
  +3\left(n_1+\tfrac{1}{2}n_2\right)R \ln \left( 1+\frac{1}{3\left(n_1+\tfrac{1}{2}n_2\right)RT_0} f(t) \right)\ .
  \end{equation}

The equilibrium point of the ODE \eqref{mec1v}--\eqref{tdv} is given by $\vec x=\vec 0$ and $\vec p=\vec 0$ and this corresponds to the state obtained in the limit $t\to \infty$, i.e. the system evolves to the equilibrium point of the ODE. In this limit the entropy tends to
\begin{equation}\label{Sequilibrium}
\begin{aligned}
&  S_{\rm eq}=\lim_{t\to \infty}S(t)=S_0+3\left(n_1+\tfrac{1}{2}n_2\right)R\\
&\times \ln \left(\frac{3\left(n_1+\tfrac{1}{2}n_2\right)RT_0+\left(\tfrac{1}{2}mv_0^2+\tfrac{1}{2}kx_0^2\right)}{3\left(n_1+\tfrac{1}{2}n_2\right)RT_0}\right) \ .
\end{aligned}
\end{equation}
Note that $\tfrac{1}{2}mv_0^2+\tfrac{1}{2}kx_0^2=-\Delta E^{\rm mec}=\Delta U$.

We remark that in this example the \emph{second law of thermostatics} is satisfied, i.e. an isolated system evolves toward an equilibrium state which is a maximum of the entropy under the condition that the energy is fixed by the initial condition. Indeed
\begin{equation}
  S=S_0+3\left(n_1+\tfrac{1}{2}n_2\right)R \ln \left( \frac{E-(\frac{1}{2}mv^2+\frac{1}{2}kx^2)}{U_0} \right)
\end{equation}
and the maximum of $S$ under the condition that \mbox{$E=\frac{1}{2}mv_0^2+\frac{1}{2}kx_0^2+U_0$} is given by the solution of \eqref{p1} and \eqref{p2}
\begin{eqnarray}
  \frac{\partial S}{\partial x}&=& -\frac{3\left(n_1+\tfrac{1}{2}n_2\right)RU_0}{ E-(\frac{1}{2}mv^2+\frac{1}{2}kx^2)}kx=0 \label{p1}\\
  \frac{\partial S}{\partial v}&=&\frac{3\left(n_1+\tfrac{1}{2}n_2\right)RU_0}{ E-(\frac{1}{2}mv^2+\frac{1}{2}kx^2)}mv=0  \label{p2}\ ,
\end{eqnarray}
i.e. $x=0$, $v=0$ and $E=U=3(n_1+\frac{1}{2}n_2)RT$.
Therefore $S_{\rm eq}$ of equation \eqref{Sequilibrium} is the maximum of $S$ under the condition $E=E_0=E(t_0)$. Moreover this maximum of the entropy gives the equilibrium state.


\section{Solid friction with $\vec F_{\rm el}=\vec 0$}

We consider a solid $\Sigma_1$ sliding on a fixed solid $\Sigma_2$ (see figure~\ref{fig2td} without the spring and $\vec p_0\not =\vec 0$). The only action of the outside on the system is the passive force maintaining fixed the solid $\Sigma_2$. We assume that the solid friction is given by \eqref{solidfr}, and that the temperatures of both systems are always equal $T_1=T_2=T$ (i.e. only one thermodynamical variable is needed to describe the state of the system), while the internal energy of $\Sigma_i$ is given by $U_i=3n_iRT_i$ ($i=1,2$). Since energy is an extensive observable $U=3(n_1+n_2)RT$. We assume that experiments have shown that $m$ the mass of $\Sigma_1$ and $\lambda$ the friction coefficient are constant.

Applying the first law of thermodynamics \eqref{q} (with $P_W^{\rm ext}=0$, $P_Q^{\rm ext}=0$) to the whole system we have
\begin{equation}\label{EC3}
  \frac{\D E}{\D t}=\frac{\D E^{\rm kin}_1}{\D t}+  \frac{\D U}{\D t}=0
\end{equation}
but $\frac{\D E^{\rm kin}_1}{\D t}=P_W^{2\to 1}=\vec F_{\rm fr}\cdot \vec v=-\lambda v$ where $v=|\vec v|$, thus
\begin{equation}\label{EC4}
  \frac{\D U}{\D t}=3(n_1+n_2)R\frac{\D T}{\D t}=\lambda v \ .
\end{equation}
Then, applying the first law of thermodynamics \eqref{q} to the two subsystems, together with $P_W^{2\to 1}=-\lambda v$ and \mbox{$P_{W}^{1\to 2}=\lambda v$}, we have
\begin{subequations}\label{eq55}
\begin{eqnarray}
    \frac{\D E_1}{\D t}=\frac{\D E^{\rm kin}_1}{\D t}+ \frac{\D U_1}{\D t}&=&P_Q^{2\to 1}-\lambda v \\
    \frac{\D E_2}{\D t}= \qquad \qquad   \frac{\D U_2}{\D t}&=&P_Q^{1\to 2}+\lambda v
\end{eqnarray}
\end{subequations}
and thus
\begin{eqnarray}
    \frac{\D U_1}{\D t}&=&P_Q^{2\to 1}\\
    \frac{\D U_2}{\D t}&=&P_Q^{1\to 2}+\lambda v\ .
\end{eqnarray}
From the energy conservation \eqref{EC3}, together with the assumption that both systems are at the same temperature and thus from \eqref{EC4}, $\frac{\D U_i}{\D t}=3n_iR\frac{\D T}{\D t}=\frac{n_i}{n_1+n_2}\lambda v$, it follows that
\begin{equation}\label{eq59}
  P_Q^{2\to 1}=-P_Q^{1\to 2}=\frac{n_1}{n_1+n_2}\lambda v \ .
\end{equation}
Therefore from \eqref{eq55} and \eqref{eq59} we obtain
\begin{eqnarray}
    \frac{\D E_1}{\D t}&=&-\frac{n_2}{n_1+n_2}\lambda v\\
    \frac{\D E_2}{\D t}&=&\frac{n_2}{n_1+n_2}\lambda v \ .
\end{eqnarray}

The subsystem $\Sigma_1$ looses $\lambda v$ kinetic energy per unit time due to the friction work power and gains $\frac{n_1}{n_1+n_2}\lambda v$ internal energy per unit of time due to the heat power received from the subsystem $\Sigma_2$. Therefore its total energy $E_1$ decreases per unit of time by $-\frac{n_2}{n_1+n_2}\lambda v$. The subsystem $\Sigma_2$ gains $\lambda v$ as work power due to the force of $\Sigma_1$ to $\Sigma_2$, but looses a quantity $\frac{n_1}{n_1+n_2}\lambda v$ as heat power to $\Sigma_1$. Therefore its total energy $E_2$ increase per unit of time by $\frac{n_2}{n_1+n_2}\lambda v$.

We now address the question of time evolution. From Newton's equation
\begin{equation}\label{newton2}
  m\frac{\D v}{\D t}=-\lambda \qquad \text{if $v>0$}
\end{equation}
we have for $v_0>0$
\begin{equation}
\begin{cases}
  v(t)=v_0-\frac{\lambda}{m}t &\qquad \text{for $0\leq t\leq \frac{mv_0}{\lambda}$} \\
  v(t)=0 &\qquad \text{for $t\geq \frac{mv_0}{\lambda}$}\ .
\end{cases}
\end{equation}

From \eqref{EC4} and the expression of $v(t)$ follow that the time evolution of temperature is given by
\begin{equation}
  T(t)=T_0+\frac{\lambda}{3(n_1+n_2)R} \left(v_0t-\frac{1}{2}\frac{\lambda}{m}t^2\right) \ ,
\end{equation}
for $0\leq t\leq \frac{mv_0}{\lambda}$, and $T(t)=T_0+\frac{1}{3(n_1+n_2)R}\frac{m}{2}v_0^2$ for $t\geq \frac{mv_0}{\lambda}$.\\
Using \eqref{entropo} we find the time evolution for the entropy, for $0\leq t\leq \frac{mv_0}{\lambda}$
\begin{equation}
  S(t)=S_0+3(n_1+n_2)R\ln \left(1+\frac{\lambda(v_0t-\tfrac{1}{2}\tfrac{\lambda}{m}t^2)}{3(n_1+n_2)RT_0} \right)\ .
\end{equation}
and for $t\geq \frac{mv_0}{\lambda}$
\begin{equation}
S(t) = S_0 + 3 (n_1 + n_2)R \ln\left(1 + \frac{\frac{1}{2}mv_0^2}{3(n_1+n_2)RT_0}  \right) \ .
\end{equation}
We observe that $S(t)$ is a strictly increasing function of time, which is the second law for the adiabatically closed system $\Sigma$.

From \eqref{entropo} and $\frac{1}{2} mv^2 + U = E = E_0$ we have
\begin{equation}
S = S_0 + 3(n_1+n_2) R \ln\left(\frac{E- \frac{1}{2}mv^2}{U_0}\right)
\end{equation}
Therefore the maximum of $S$ under the condition that $E = E_0$ is fixed gives $v =0$ and $S_{\max}$ is thus equal to the value of the entropy at equilibrium. Again we recover in this example the second law of thermostatics, but this law does not give the equilibrium position: to obtain the equilibrium position we have to solve the time evolution \eqref{newton2} together with $\frac{\D x}{\D t}=v$.


\section{Solid friction with $\vec F_{\rm el}\not =\vec 0$}\label{attrito}

We consider the situation used in section~\ref{sect3} to build the thermodynamic laws (see figure~\ref{fig2td}), where one thermodynamical variable $S$ is sufficient to describe the state, and we assume that $m$, $\lambda$, and $k$ are constant.
We shall consider that the internal energy of both solids is given by $U_i = 3 n_iRT$, with $n_i$ the number of moles in solid $i$ ($i= 1, 2$) and, since energy is an extensive observable, for the total system $\Sigma$ we thus have $U=3(n_1+n_2)RT$.

We have to solve the ODE giving the time evolution of the state $(\vec x,\vec p,S)$ when the only action from the outside is the passive force keeping fixed the solid $\Sigma_2$ (equations \eqref{mec1}--\eqref{td} with $\vec F_{\rm ext}=\vec 0$). We suppose that the initial mechanical state is given by $(\vec x_0,\vec p_0)$ with $x_0>\frac{F_{\rm fr,max}}{k}$ and $p_0=0$. Explicitly \eqref{mec2} writes as
\begin{equation}\label{C1}
\frac{\D \vec p}{\D t}=\begin{cases}
  \vec F_{\rm el} +\vec F_{\rm fr} \quad &\text{if } \vec p\not =\vec 0 \text{ or } \vec p=\vec 0 \text{ and } |\vec F_{\rm el}|>F_{\rm fr,max} \\
  0 \quad & \text{ otherwise}
\end{cases}\
\end{equation}
with $\vec F_{\rm el}=-k\vec x$ and $\vec F_{\rm fr}=-\frac{\lambda}{m} \hat p$ if $\vec p\not =\vec 0$.
Note that because of \eqref{C1} it is not possible to find the equilibrium state without integrating the system of ODE. Therefore to find the equilibrium position $x_{\rm eq}$ we have to solve explicitly the time evolution, which is straightforward and gives in general $x_{\rm eq}\not =0$. We then investigate the time evolution of the entropy.

Since the evolution was derived from energy conservation we obtain immediately the time evolution for the temperature. Indeed from $E^{\rm mec}(t)+U(t)=E^{\rm mec}_0+U_0$, i.e.
\begin{equation}
\frac{1}{2} mv(t)^2 + \frac{1}{2} kx(t)^2 + U(t) = \frac{1}{2} k x_0^2 +U_0
\end{equation}
and $U = 3(n_1+n_2) RT$, we have
\begin{equation}\label{Tmas}
T(t) = T_0 + \frac{1}{3(n_1+n_2)R} \frac{1}{2} k\left(x_0^2 - x(t)^2 -\frac{m}{k} v(t)^2\right)\ .
\end{equation}
The time evolution for the entropy $S(t)$ is then given by \eqref{entropo} and \eqref{Tmas}, i.e.
\begin{equation}
S(t) = S_0 + 3(n_1+n_2) R \ln\left(\frac{E_0- \left(\frac{1}{2}mv(t)^2-\frac{1}{2}kx(t)^2\right)}{U_0}\right)
\end{equation}
Once more this example shows that the entropy of the isolated system is an increasing function of time as postulated in the second law of thermodynamics.

At equilibrium we will have
\begin{equation}
  T_{\rm eq}=T_0+\frac{1}{3(n_1+n_2)R}\frac{k}{2}(x_0^2-x_{\rm eq}^2)\, ,
\end{equation}
and
\begin{equation}\label{equilbs}
  S_{\rm eq}=S_0
  +3(n_1+n_2)R
     \ln \left(1+\frac{\frac{1}{2}k}{3(n_1+n_2)RT_0}\left(x_0^2-x_{\rm eq}^2\right)\right) \ .
\end{equation}
Observe that $\frac{k}{2}\left(x_0^2-x_{\rm eq}^2\right)$  is the decrease of potential energy $-\Delta E^{\rm pot}=\Delta U$.
Note that in this example the second law of thermostatics stated in \ref{secA} is not satisfied.
Indeed the maximum of the entropy under the condition that $E = E_0$ is fixed gives $v=0$ and $x =0$ which is not the equilibrium state as one can see by solving explicitly the time evolution equation.
Once more solid friction give examples where to find the equilibrium state we have to solve explicitly the time evolution and it is not possible to apply a maximum entropy condition. This situation is analogous to the well known ``adiabatic piston problem'' \cite{CG3,HC}.


\section{A first generalization of section~\ref{sect4}}\label{appD}

In all previous examples we have assumed that the state of the whole system was described by only one thermodynamical variable, which implies in particular that the temperatures of both subsystems are equal at all time. We now drop this condition and we assume that both solids are described by one thermodynamical variable, which is either the temperature $T_1$, $T_2$, or the internal energy $U_1$, $U_2$, or the entropy $S_1$, $S_2$. The relation between the entropy and the temperature is given for each solid by \eqref{entropo}.

We consider first the system described in section~\ref{sect4} in the case where $\vec F^{\rm ext}=\vec 0$ and $\vec \omega=\vec 0$. We have
\begin{equation}
  \frac{\D U}{\D t}=0
\end{equation}
for the whole system, and from \eqref{tempass} and \eqref{q} for the two subsystems
\begin{eqnarray}
\frac{\D U_1}{\D t}&=&T_1\frac{\D S_1}{\D t}=P_Q^{2\to 1} \\
\frac{\D U_2}{\D t}&=&T_2\frac{\D S_2}{\D t}=P_Q^{1\to 2}
\end{eqnarray}
which implies $P_Q^{2\to 1}+P_Q^{2\to 1}=0$.\\ Since entropy is an extensive observable \mbox{$S=S_1+S_2$}, we thus obtain
\begin{equation}
  \frac{\D S}{\D t}=I=\left(\frac{1}{T_1}-\frac{1}{T_2}\right) P_Q^{2\to 1}\ .
\end{equation}
Using the second law of thermodynamics \eqref{secondalegge}, and the fact that the absolute temperature is positive, we are led to conclude that there exists a positive state function $\kappa=\kappa(T_1,T_2)$, called heat conductivity, such that
\begin{equation}\label{80}
  P_Q^{2\to 1}=\kappa(T_2-T_1)\ .
\end{equation}
We thus have
\begin{eqnarray}
3n_1R\frac{\D T_1}{\D t}&=&\kappa (T_2-T_1) \\
3n_2R\frac{\D T_2}{\D t}&=&-\kappa (T_2-T_1)\ .
\end{eqnarray}

\noindent Since experiment shows that $\kappa$ depends very slightly on the temperatures, assuming $\kappa$ to be constant we find
\begin{equation}
  T_2(t)-T_1(t)=(T_{2,0}-T_{1,0})\text{e}^{-At}\ ,
\end{equation}
and then the time evolution for the temperatures
\begin{eqnarray}
T_1(t)&=& T_{1,0}+\frac{n_2}{n_1+n_2}(T_{2,0}-T_{1,0})\left(1-\text{e}^{-At}\right)\\
T_2(t)&=&T_{2,0}-\frac{n_1}{n_1+n_2}(T_{2,0}-T_{1,0})\left(1-\text{e}^{-At}\right)
\end{eqnarray}
where
\begin{equation}\label{A}
  A=\left(\frac{1}{n_1}+\frac{1}{n_2}\right)\frac{\kappa}{3R}\ .
\end{equation}
Finally, in the limit $t\to \infty$ we obtain, as expected, the thermal equilibrium i.e. $T_1=T_2$. In this case the second law of thermostatics is satisfied.

Consider finally the example of section~\ref{sect4} without the assumption that both solids have the same temperature. Since the mass is constant we now have
\begin{eqnarray}
\frac{\D U_1}{\D t}&=&T_1\frac{\D S_1}{\D t}=P_Q^{2\to 1} \\
\frac{\D U_2}{\D t}&=&T_2\frac{\D S_2}{\D t}=\lambda \omega + P_Q^{1\to 2}
\end{eqnarray}
and $P_Q^{1\to 2}=-P_Q^{2\to 1}$. That is
\begin{eqnarray}
\frac{\D S_1}{\D t}&=&\frac{1}{T_1}P_Q^{2\to 1} \label{Sins}\\
\frac{\D S_2}{\D t}&=&\frac{1}{T_2}\lambda \omega+ \frac{1}{T_2}P_Q^{1\to 2} \label{Sins2}
\end{eqnarray}
and, since entropy is an extensive observable,
\begin{equation}\label{Stot}
  \frac{\D S}{\D t}=I=\frac{1}{T_2}\lambda \omega+\left(\frac{1}{T_1}-\frac{1}{T_2}\right) P_Q^{2\to 1}
\end{equation}
where $I$ has to be a non negative state function according to the second law \eqref{secondalegge}.

\noindent From \eqref{55} and \eqref{80} we are led to introduce the assumption for
\begin{equation}
  P_Q^{2\to 1}=C\lambda \omega + \kappa(T_2-T_1)
\end{equation}
where $C$ is a constant defined from experiments. The irreversibility state function is thus given by
\begin{equation}
  I=\frac{1}{T_1}C\lambda \omega +\frac{1}{T_2}(1-C)\lambda \omega +\frac{\kappa}{T_1T_2}(T_2-T_1)^2
\end{equation}
and the non negative condition of $I$ implies that \mbox{$C\in[0,1]$}. (In \eqref{55} $C=\frac{n_1}{n_1+n_2}$ but this will imply that the temperatures remain equal if they were initially equal.)
The equations \eqref{Sins}, \eqref{Sins2} and \eqref{Stot} are the expression of the second law of thermodynamics \eqref{secondalegge} applied to the subsystems $\Sigma_1$, $\Sigma_2$ and to the whole system $\Sigma$ which is adiabatically closed.

\noindent We now have
\begin{eqnarray}
\frac{\D T_1}{\D t}&=&\frac{C}{3n_1R}\lambda \omega +\frac{\kappa}{3n_1R} (T_2-T_1) \\
\frac{\D T_2}{\D t}&=&\frac{1-C}{3n_2R}\lambda \omega -\frac{\kappa}{3n_2R} (T_2-T_1) \ .
\end{eqnarray}
Assuming that the angular velocity $\omega$ and the heat conductivity $\kappa$ are constants, we obtain
\begin{equation}
  T_2(t)-T_1(t)=B+\left[(T_{2,0}-T_{1,0})-B\right]\text{e}^{-At}
\end{equation}
with $A$ given by \eqref{A} and
\begin{equation}
  B=\frac{n_1-C(n_1+n_2)}{\kappa (n_1+n_2)}\lambda \omega\ .
\end{equation}
In conclusion for $t\to \infty$
\begin{equation}
  T_2(t)-T_1(t)\to B
\end{equation}
and
\begin{equation}
  P_Q^{2\to 1} \to C\lambda \omega +\kappa B=\frac{n_1}{n_1+n_2}\lambda \omega
\end{equation}
i.e. the heat transferred per unit time from $\Sigma_2$ to $\Sigma_1$ does not depend on the unknown constant $C$.




\end{document}